# An Optical Test Bench for the Precision Characterization of Absolute Quantum Efficiency for the TESS CCD Detectors

A. Krishnamurthy, J. Villasenor, S. Kissel, G. Ricker, R. Vanderspek
MIT Kavli Institute for Astrophysics and Space Research, Massachusetts Institute of Technology, Cambridge, USA

## ABSTRACT

The Transiting Exoplanet Survey Satellite (TESS) will search for planets transiting bright stars with Ic ≲13. TESS has been selected by NASA for launch in 2018 as an Astrophysics Explorer mission, and is expected to discover a thousand or more planets that are smaller in size than Neptune. TESS will employ four wide-field optical charge-coupled device (CCD) cameras with a band-pass of 650 nm – 1050 nm to detect temporary drops in brightness of stars due to planetary transits. The 1050 nm limit is set by the quantum efficiency (QE) of the CCDs. The detector assembly consists of four back-illuminated MIT Lincoln Laboratory CCID-80 devices. Each CCID-80 device consists of 2048x2048 imaging array and 2048x2048 frame store regions. Very precise on-ground calibration and characterization of CCD detectors will significantly assist in the analysis of the science data obtained in space. The characterization of the absolute QE of the CCD detectors is a crucial part of the characterization process because QE affects the performance of the CCD significantly over the redder wavelengths at which TESS will be operating. An optical test bench with significantly high photometric stability has been developed to perform precise QE measurements. The design of the test setup along with key hardware, methodology, and results from the test campaign are presented.

## 1. INTRODUCTION

The Transiting Exoplanet Survey Satellite (TESS) [Ricker et al., 2014] is an Astrophysics Explorer mission selected by NASA for launch in 2018 to search for planets transiting bright stars with Ic ≲13. TESS will employ four identical wide-field optical CCD cameras with a band-pass of 650 nm – 1050 nm to perform differential time-series photometry by monitoring at least 200,000 main sequence stars. Each camera consists of an f/1.4 custom lens assembly, and a CCD detector assembly, which consists of four deep-depletion back-illuminated MIT Lincoln Lab CCID-80 devices [Suntharalingam et al., 2016] with associated electronics. The electronics consist of three compact double-sided printed circuit boards, each 12 cm in diameter. The detectors are designed for enhanced sensitivity to the redder wavelengths because it is easier to detect small planets around small red stars. The upper limit of the band-pass cutoff at 1050 nm is driven by the quantum-efficiency curve of the detectors. A higher QE over the red wavelengths will yield a higher photon count, and thus higher planetary detections [Sullivan et al., 2015].

Precise characterization of the absolute quantum efficiency of the CCD detectors is essential to enable accurate modeling of the projected photon count as well as spectral weighting of the incoming photons. Previous works by Poletto et al., 1997 and Groom et al., 2006 have established methodologies to characterize the detector QE with an uncertainty of approximately 3-6% and 3.6% respectively, by comparing the signal produced by the CCD to the incoming photon current measured using a calibrated photodiode. This paper presents an approach to further lower the uncertainty in the measurements to within 2-2.5% by constructing a precision test bench with careful light source selection and stabilization, filter selection, reference detector calibration and precise gain measurements.

## 2. PRECISION TEST BENCH

The test setup [Krishnamurthy et al., 2016], illustrated in Figure 1, consists of a vacuum chamber with a single MIT Lincoln Lab CCID-80 device mounted on a cold plate that is maintained at the operating temperature of -70°C to reduce the dark current to a negligible level. A calibrated reference photodiode is mounted next to the CCD and maintained at the calibration temperature of 25°C. Band-pass filters over the range of 600 nm – 1064 nm with 10-nm bandwidth are used for wavelength selection.

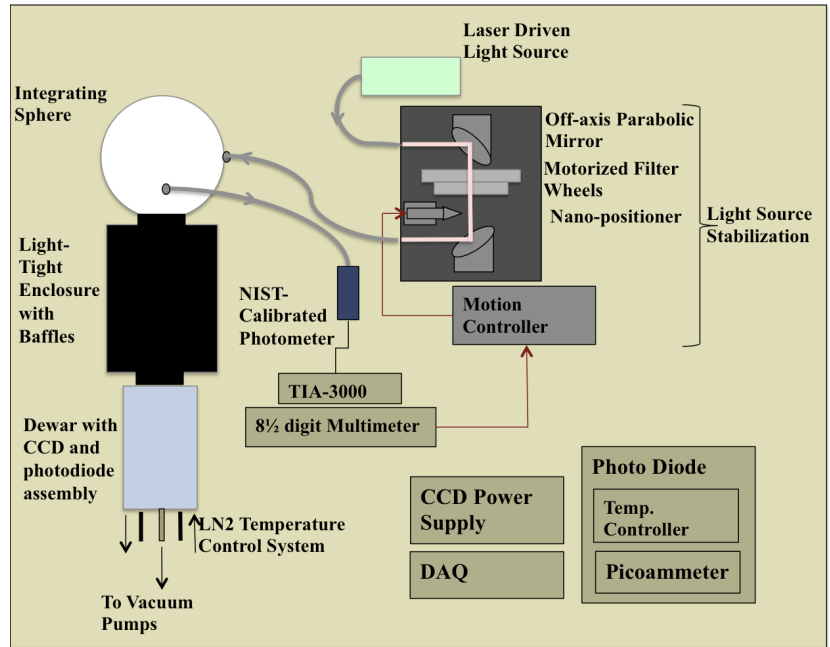

Figure 1: Precision optical test bench for the characterization of absolute quantum efficiency.

A very stable laser-driven light source (LDLS) is integrated with the Super Stable Source (SSS) stabilization unit, a patented development by the Characterising ExOPlanetS (CHEOPS) Team at the University of Geneva [Francois Wildi, 2015], to control variations of the light source down to a few parts-per-million when averaged over 60 s. Light from the stabilization unit enters a 20-inch integrating sphere to produce near-uniform diffuse illumination on the CCD and on the calibrated reference photodiode simultaneously. A light-tight enclosure is installed between the integrating sphere exit port and the CCD to obtain near-uniform illumination of the CCD and to eliminate light leaks. A set of baffles inside the black interior is used to prevent stray light and secondary reflections from entering the vacuum chamber.

**Uniform Optical Illumination**

The Energetiq Laser-Driven Light Source (LDLS) technology consists of a continuous wave laser plasma discharge with plasma size of ~100 $\mu$m and a broad spectral spectrum of 190 nm – 2400 nm. The CW laser directly heats Xenon plasma created in a fused silica bulb using a traditional arc light igniter, to a very high blackbody temperature of 10,000 K. The light source has excellent power stability with instantaneous variation of about 1-3%. The electrodeless operation contributes to the long lamp life of over 9000 hours.

Since we are focused on characterizing the absolute quantum efficiency of the CCID-80 device in the 650 nm – 1050 nm spectral range, a 12-position Thorlabs motorized filter wheel is used to perform careful filter selection. It contains twelve hard coated band-pass filters from Edmund Optics that provide deeper blocking and higher transmission (>90%) compared to traditional coated filters. Eleven of them have a bandwidth of 10-nm and are over the range of 650 nm – 1064 while one filter at 1000 nm has a 25-nm bandwidth. A 6-position motorized filter wheel with 1-inch absorptive neutral density filters are used to control the intensity of the light entering the integrating sphere.

A 20-inch custom-built integrating sphere from LabSphere is used is to produce uniform diffuse illumination over the CCD and calibrated photodiode assembly in the vacuum chamber. The integrating sphere consists of a hollow spherical cavity with a diffuse white reflective coating called Spectraflect with

98% reflectance over the 650 nm – 1050 nm spectral range. It has one 1-inch inlet port and one 4-inch outlet port placed at right angles to each other to make sure each ray of light is reflected at least once before exiting the output port. Light rays incident on any point on the inner surface are, by multiple scattering reflections, distributed equally to all other points thus preserving power and destroying spatial information producing diffuse light at the outlet. There is a 1-inch diagnostic port above the output port to make photometric measurements of the light inside the sphere that is used by the light stabilization unit to control the flux variations of the LDLS.

**Light Stabilization**

The light stabilization unit is a patented development by the Characterising ExOPlanetS (CHEOPS) Team at the University of Geneva [Wildi et al., 2015]. It is integrated with the test up in order to stabilize the Energetiz Laser-Driven Light Source. The stabilization unit consists of a fibered input and fibered output with a precision light control system implemented between the two. There is a collimator-decollimator assembly consisting of two off-axis parabolic mirrors, two motorized filter wheels for filter selection and intensity control, and a knife-edge attenuator. The knife-edge attenuator has a conical end towards the beam and is mounted on a precision positioner with a maximum reachable velocity of 4.5 mm/s to occult the beam and balance the flux variations of the LDLS. A feedback loop with a NIST-calibrated precision photometer is mounted on one of the output ports of the integrating sphere to control the movement of the attenuator into the beam such that the variation of flux at the output fiber is minimum. The measurement uncertainty of the photometer system is about 5 ppm at 15 Hz and 24-hour accuracy below 1 ppm.

The closed loop stabilization stage operates in two modes; a slow mode that commands the attenuator and waits for it to reach the required position before measuring the next point, and a continuous mode where the photometer system measures the flux without waiting for the attenuator to reach the point previously commanded. The slow mode yields a stability of 4.95 ppm and the continuous mode yields a stability of 3.57 ppm when averaged over 60 s. The slow mode was used to stabilize the LDLS at each wavelength before recording the QE measurements.

**Dewar Assembly**

The dewar assembly consists of a custom-built stainless steel chamber that is 14 inch in length and 10 inch in diameter with a 7.37 inch optical quartz window. The CCD and calibrated photodiode assembly are mounted on a ¼" thick 6061 Aluminum cold plate attached to the front end of an annular liquid Nitrogen reservoir that maintains the CCD at the operating temperature of -70 ℃. The CCD electronics are placed behind the CCD and photodiode assembly. A black anodized plate is used to mask the CCD and calibrated photodiode assembly with two apertures over the CCD imaging array area and the photodiode.

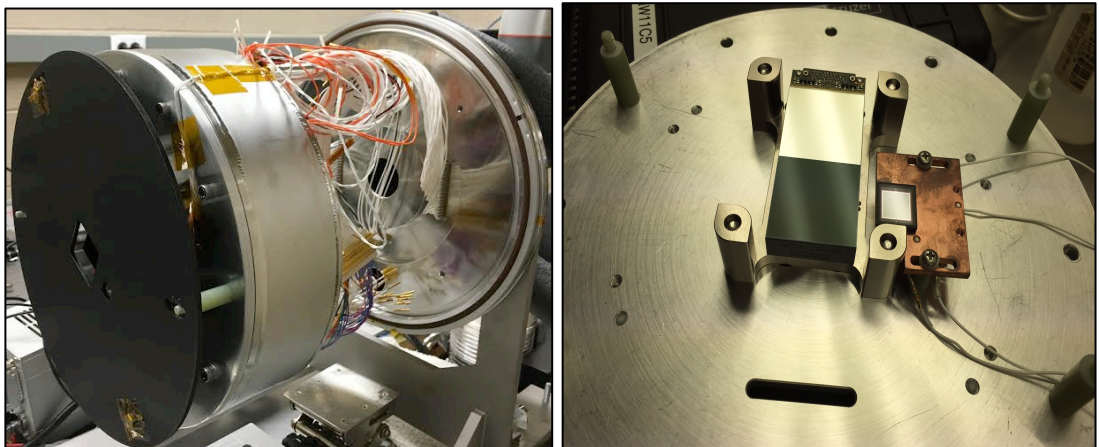

Figure 2: [Left] Open vacuum chamber showing the black anodized mask plate mounted on the cold plate using insulating standoffs, and the LN2 reservoir behind the cold plate that maintains it at the operating temperature of -70

°C. [Right] CCD and calibrated photodiode assembly mounted on the cold plate. The photodiode assembly is mounted using insulating standoffs, and has a heater to maintain the photodiode at an operating temperature of 25 °C.

**Calibrated Photodiode**

The CCD and the calibrated reference photodiode are placed next to each other on the cold plate, as shown in Figure 2, within a circular area of 4-inch diameter such that they are in the same focal plane and intercept the incoming light from the integrating sphere simultaneously. The calibrated reference photodiode model used for the QE measurements is the Hamamatsu S1337-1010BQ, a Silicon photodiode with a Quartz window, and a 10 mm x 10 mm photosensitive area. The precision photometer in the light stabilization setup is used to cross-calibrate the reference photodiode to improve the accuracy of the measurements. The CCD is maintained at the operating temperature of -70°C while the reference photodiode is maintained at the calibration temperature of 25°C using a 5W heater mounted on the bottom of the copper plate. The sensitivity measurements of the photodiode were performed by Hamamatsu and the calibration curve is shown in Figure 3.

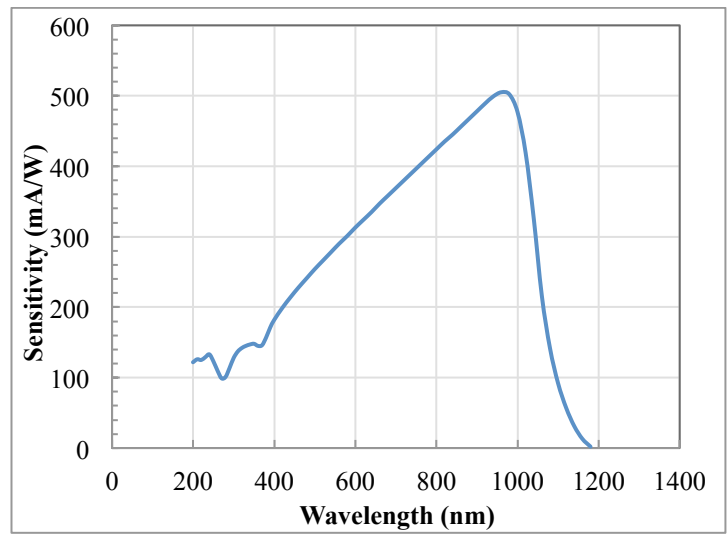

Figure 3: Calibrated photodiode sensitivity [mA/W] plotted against wavelength [nm].

**CCD Readout**

Each CCID-80 device consists of imaging array and frame store regions. The imaging array as seen in Figure 4, consists of 2048 rows by 2048 columns plus one set of buffer rows each at the top and bottom. The buffer rows do not count the signal consistently, and hence not used in our output signal calculations. The image array and the buffer rows are transferred to the frame store. The frame store region also measures 2048 x 2048, and has two sets of buffer rows, one at the top and one at the bottom, adding up to 2068 rows by 2048 columns. There are also 10 buffer columns on either side of the imaging array and frame-store regions that are not clocked into the serial register.

There are 10 smear rows that are created during parallel clocking, and are transferred from the image array to the frame store, and 10 virtual rows that are clocked during the frame-store readout that are not exposed to any illumination, and hence can be used as a measure of the dark current. The transfer of rows into the frame-store region is performed at a total read out time of 19.95 ms. The pixel outputs are then transferred to the serial register, and read out individually.

The CCD readout has four output registers: A, B, C, and D. The readout of the pixels from sectors B and D occur in the opposite direction from those through outputs A and C. This is accounted for in the generation of the FITS images. For the QE measurements, we use a single flight-grade engineering CCD with pre-flight electronics, and the output from the CCD is obtained in a FITS-format file with the pixels reconstructed such that the four outputs are placed adjacent to each other.

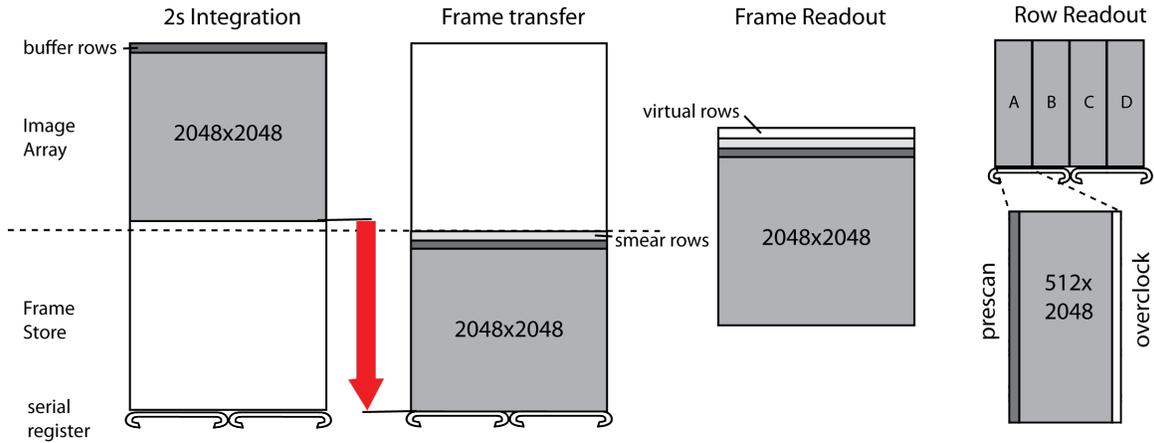

Figure 4: Schematic of the CCID-80 device showing the imaging array and frame-store regions, and readout directions for the four sectors A, B, C, and D. The imaging array is has 2048 x 2048 pixels with 512 columns allocated to each of the four sectors A, B, C, and D. In addition, there are 10 virtual rows, 10 smear rows and 10 buffer rows, and 11 virtual underclock and 11 virtual overclock columns. [Thayer et al., 2016]

## 3. ABSOLUTE QUANTUM EFFICIENCY

The absolute quantum efficiency, QE is given by ratio of the output signal to the incoming photon current.

$$QE = \left(\frac{(Output\ signal\ per\ pixel) * Gain}{Incoming\ photons\ per\ pixel}\right) \quad (1)$$

Where output signal per pixel is expressed in [ADU] and gain conversion factor is expressed in [e$^-$/ADU]. Measurement of gain is discussed in the following section. To calculate the output signal, we take dark frames, and 20 2-second images with uniform illumination at each wavelength. We then subtract the dark frames from the images and stack the 20 images together to calculate the median output signal per pixel.

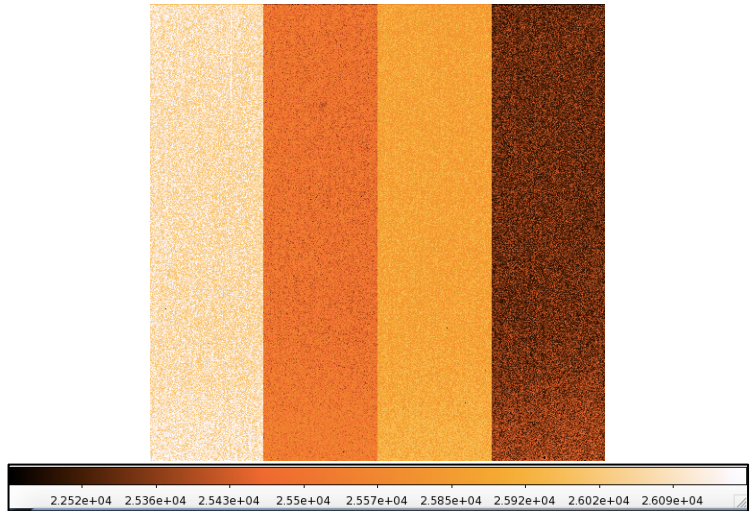

Figure 5: FITS image at 750 nm. For wavelengths shorter than 800 nm, the CCD has a uniform spatial response, and the bright lines due to the straps and temperature sensors are not seen.

One important factor in determining what parts of the CCD to use in calculating the output signal is driven by the presence of aluminum straps and temperature sensors underneath the surface of the CCD that reflects light at certain wavelengths. For wavelengths shorter than 800 nm, the CCD has a uniform spatial response, as seen in Figure 5, but at longer wavelengths between 825 – 1050 nm, bright lines due to the straps and temperature sensors are observed, as seen in Figure 6. The increase in signal is between 0.5% - 14% as we sample across different wavelengths between 825 nm – 1050 nm. The absolute QE measured at the straps is up to 4% greater than the rest of the CCD. So, for the purpose of absolute QE measurements for the detector, we sample in between the straps so that the calculation of absolute QE remains consistent across different wavelengths. In Figure 6, we also observe the phenomenon of fringing that is caused by interference of light at the boundaries of the epoxy layer. This feature is prominently observed between 905 nm – 1000 nm due to the longer wavelengths penetrating the features in the underlying structure.

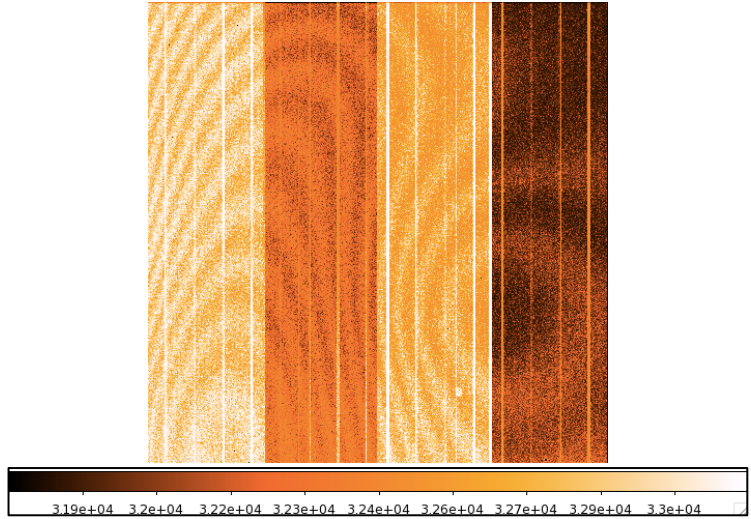

Figure 6: FITS image showing bright lines and interference pattern at 905 nm. The bright lines are a result of reflection from the aluminum straps and temperature sensors underneath the surface, and the fringing is caused by the interference due to the variation in the thickness of the epoxy bonding of the silicon to the substrate. We sample in between the straps so that the calculation of absolute QE remains consistent across different wavelengths.

Incoming photons per pixel can be calculated by multiplying the number of photons with the area of the pixel and the integration time.

$$\text{Incoming photons per pixel} = N * t \quad (2)$$

where t = exposure time = 2 s.
The number of photons per second, $N$ [photons/s] is given by

$$N = \left(\frac{P_{PD}}{E_{photon}}\right) \quad (3)$$

where $P_{PD}$ is the total incident power of the incoming photons, and $E_{photon}$ is the energy per photon.

Incident light power $P_{PD}$ [W] incident on the photodiode is given by

$$P_{PD} = \left(\frac{I}{S}\right) \quad (4)$$

where $I$ is the photocurrent [A] of the Hamamatsu photodiode, $S$ is the sensitivity [A/W] of the photodiode at a given wavelength. Sensitivity of the photodiode is given by the calibration curve, as shown in Figure 3, obtained from the manufacturer of the photodiode.

Energy per photon, $E_{photon}$ [J], is given by

$$E_{photon} = \left(\frac{hc}{\lambda}\right) \quad (5)$$

where h = 6.62607004 × 10$^{-34}$ [m² kg / s], c is the speed of light [m/s], and $\lambda$ is the wavelength of light [nm].

## 4. RESULTS AND DISCUSSION

**Gain Measurements:**
The gain is the conversion factor between the electrons collected in the CCD and the Analog-to-Digital readout Units (ADU). Gain is dependent on the temperature of the CCD, and is measured using the $Cd^{109}$ $K_\alpha$ and $K_\beta$ peaks. Using the known energies of the X-ray peaks ($K_\alpha$=22.1 keV, $K_\beta$ = 25.0 keV), and the conversion of eV/e- for Silicon [Groom et al., 2006], we fit a line to the peaks whose slope gives the gain conversion factor [Schloze et al., 1996]. The gain measured is 6.99 ± 0.01 e-/ADU [Krishnamurthy et al., 2016]. The gain measurements are verified by calculating the mean variance from the photon transfer measurements.

**QE Measurements**

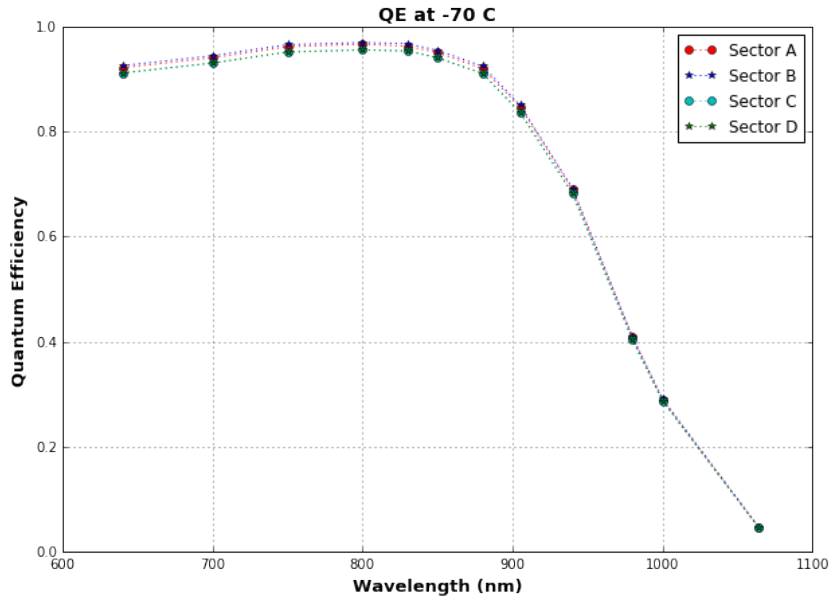

Figure 7: Quantum efficiency measurement for the TESS flight-grade 100-micron thick CCID-80 device with pre-flight electronics.

The operating temperature of the CCD was -70 °C and the exposure time was 2 s. Figure 7 shows the absolute QE measurements obtained from the precision optical test setup, plotted individually for the four output sectors A, B, C, and D. Sectors A and B have a slightly higher QE than sectors C and D at the shorter wavelengths, while the four sectors display identical quantum efficiency at the redder end of the spectrum. Sources of error like out-of-bandpass leakage from filters, light leaks, second-order reflections, and noise in the reference photodiode measurements were found to be negligible.

## Temperature Dependence

The QE measurement along with the lens throughput is used to calculate the spectral weightings, and thus the normalized photo counts for stellar spectra. Hence, it is important to characterize the performance of the detectors at different operating temperatures, in order to study the optimal operating temperature for specific stellar types. At -50 °C, there is an increase in QE of about 4% at 1000 nm, and at -25 °C, QE at 1000 nm is about 39%. However, dark current increases substantially at higher temperatures, negating the gain in QE.

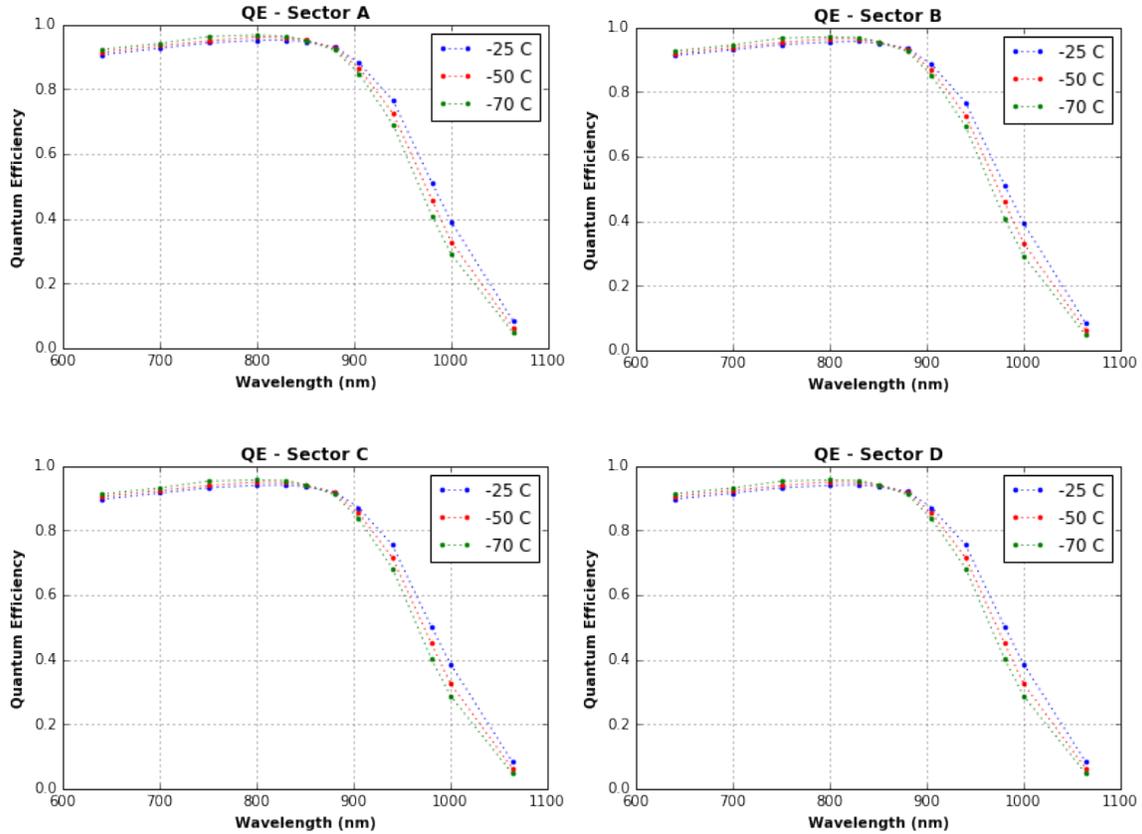

Figure 8: Temperature dependence of the absolute QE measurements are plotted individually for sectors A, B, C, and D of a 100-micron thick CCID-80 device. The operating temperature of the CCD was maintained at -25°C, -50°C and -70°C. The exposure time was 2 seconds.

## ACKNOWLEDGEMENTS

This work is supported by NASA under the contract number NNG14FC03C.